# Automatic Music Transcription using Convolutional Neural Networks and Constant-Q transform

Yohannis Telila[1,*], Tommaso Cucinotta[2] and Davide Bacciu[3]

[1]*Università di Pisa, Area CNR, Pisa, 56100, Italy*
[2]*Scuola Superiore Sant'Anna, Area CNR, Pisa, Italy*
[3]*Università di Pisa, Largo B. Pontecorvo 3, Pisa, 56127, Italy*

**Abstract**

Automatic music transcription (AMT) is the problem of analyzing an audio recording of a musical piece and detecting notes that are being played. AMT is a challenging problem, particularly when it comes to polyphonic music. The goal of AMT is to produce a score representation of a music piece, by analyzing a sound signal containing multiple notes played simultaneously. In this work, we design a processing pipeline that can transform classical piano audio files in .wav format into a music score representation. The features from the audio signals are extracted using the constant-Q transform, and the resulting coefficients are used as an input to the convolutional neural network (CNN) model.

**Keywords**
Automatic Music Transcription, Constant-Q transform, Deep Learning, Multi-pitch estimation,

## 1. Introduction

In music, transcription refers to the process of analyzing an acoustic musical signal and writing down the musical information such as the pitch, onset time, duration, and the type of each sound that occurs in it. To visually represent this information, musical notations are often used. In a computational music transcription system, a MIDI music representation format is often appropriate.

There are various reasons why having music represented in a transcribed format is important. Transcribed music is of great help for musicians, since they do not have to memorize the whole music composition to play it. Additionally, music that has been transcribed can be shared and made accessible to other musicians.

The term "Automatic music transcription" was first introduced by the researchers James A. Moorer, Martin Piszczalski, and Bernard Galler in their audio research in 1977. With their audio engineering research and knowledge, they believed that a computer could be programmed to detect chord patterns and rhythmic accents in a digital recording of music [1].

AMT is a challenging problem, particularly when it comes to polyphonic music. Several factors make AMT challenging. Polyphonic music is a complex mixture of numerous sources, such as instruments and vocals with different onset, pitch, and volume characteristics. Inferring individual musical attribute (e.g., pitch) from the mixture within the signal is a highly challenging problem. In multi-pitch estimation, the individual harmonics from the notes that make up the chord can cancel with each other. This results in a decrease or complete elimination of certain frequencies, or in some cases these harmonics can add up with each other, leading to an increase in the overall amplitude of the frequencies in the combined signal. As a result, training a model for multi-pitch estimation becomes challenging. Another key challenge for AMT is the lack of a general and sizable dataset.

## 2. Related works

While monophonic music transcription has been extensively studied and solutions have been proposed, the automatic transcription of polyphonic music remains a complex and unsolved problem. To tackle this problem, various approaches have been proposed, with two algorithm families having dominated during the last decade: Non-negative matrix factorization (NMF) and Neural networks (NNs). Both methods have been used for various tasks, including speech, image and natural language processing and recommender systems.

### 2.1. Non-negative Matrix Factorization for AMT

Non-negative matrix factorization (NMF) is a numerical algorithm that decomposes a non-negative matrix into two low rank approximation non-negative matrices and can be used for extracting meaningful patterns [2].

The first implementations of NMF in musical applications were reported in [2] and [3], where it was used for transcribing polyphonic music for separating various sound sources. All approaches used learn representations of audio segments that correspond to musical events. The work by Arshia Cont [4] introduces a modified NMF algorithm with sparseness constraints to predict pitch in real-time, so that the number of non-zero elements in the solution are minimized. This modification allows for an efficient and better detection of the pitch in real-time scenarios by reducing the number of non-zero elements in the NMF factor matrices. Additionally, the sparseness constraint ensures that only relevant information is extracted, leading to a more precise and concise representation of the audio signal since NMF usually produces a sparse representation of the data.

### 2.2. Neural Networks for AMT

Recent advancements in deep learning technology in pattern recognition and audio processing sparked the interest of researchers in the field of Music information retrieval (MIR). There are several related works that apply neural networks in AMT.

In [5], Zalani and Mittal proposed a polyphonic music transcription system based on deep learning. They trained 88 binary classifiers capable of transcribing notes of polyphonic music. Each classifier detects the presence of a single note in the music at each time step. The authors proposed a Recurrent Neural Network (RNN) [6] stacked with Restricted Boltzmann Machines (RBM) [7] for unsupervised feature learning and trained separate 88 support vector machines (SVM) [8] responsible for transcribing single notes. To enhance the accuracy of their results, a Hidden Markov Model (HMM) [9] was also implemented.

In [10], Sigtia et al. proposed an end-to-end approach for transcribing a piece of piano music into music score. The authors mentioned that their architecture is the combination of both an acoustic model and a music language model. They experimented with three different neural networks for the acoustic model, namely an artificial neural network, an RNN, and CNN, all of which were trained to recognize CQT coefficients. For the music language model, the author experimented with three types of neural networks - the generative RNN, neural autoregressive distribution estimator (NADE) proposed by Larochelle and Murray et al. [11], and RNN-NDE for the music language model. The results demonstrate that the CNN outperformed the other architectures in all evaluation metrics.

Kelz et al. [12] conducted an extensive analysis of only using an acoustic model for transcription in order to explore the limitations of simple approaches in the automatic music transcription task, including suitable input representations and they performed hyper-parameter tuning to improve the performance.

The authors also adopted the MAPS dataset [13], which provides MIDI-aligned recordings of numerous classical music pieces. The authors conducted two sets of experiments, which focused on types of input representations and hyper-parameter search to achieve the best possible results. The four types of input representations compared in the study were spectrograms with linearly spaced bins **S**, spectrograms with logarithmically spaced bins **LS**, spectrograms with logarithmically spaced bins, and logarithmically scaled magnitude **LM** as well as the CQT. The authors' analysis showed that the **S** type was the best data representation for the perceptron network (Logistic regression), whereas the **LM** type was the best for the shallow net. The authors mentioned that the lower performance of the CQT was quite unexpected and suggested further investigation.

The authors used practical training recommendations with Stochastic Gradient Descent (SGD) and found that adaptive learning rate strategies, such as Adam, improved the performance to some extent but did not eliminate the need for tuning the initial learning rate and its schedule.

The current state of the art method for general purpose piano transcription was proposed by Google Brain, by Hawthorne et al. in [14]. The approach uses two distinct architectures to detect note onsets, which is a convolutional layer and its output is fed to a second network which is a bidirectional LSTM with 128 units in both directions connected to a fully connected sigmoid layer with 88 outputs to represent the 88 piano keys. The paper also uses the MAPS dataset.

## 3. Proposed Approach

For AMT, a holistic approach is commonly used, which involves using a single model to identify and detect all the musical notes in a given piece of music (see Fig. 1).

### 3.1. CNN architecture

In this work, an architecture is proposed for the AMT task involving the use of CNNs and the Constant-Q transform (CQT). The features from the audio signal are extracted using the CQT method, and the resulting CQT coefficients are used as an input to the CNN model. CNNs have become increasingly popular, particularly in the computer vision field [15]. In computer vision, an input image is fed into the CNN and passed through multiple layers of different nature, including convolutional, pooling, and activation layers, where non-linearities may be applied. The convolutional layers are made up of various filters, each capable of learning higher-level image features. In the context of AMT, note detection can be approached similarly to image recognition, as it is possible to create time-frequency representations of audio that resemble images.

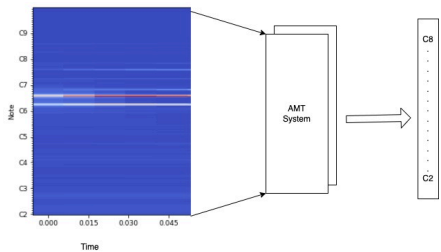

**Figure 1:** Holistic approach of AMT: Involves analyzing the complete spectrum of a polyphonic audio signal and predicting all possible notes in a single frame, capturing complex note relationships and generating a possibly accurate transcription of the musical piece.

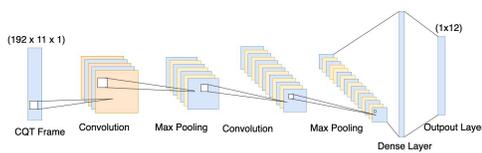

**Figure 2:** CNN architecture of AMT

## 4. Data Preparation

As described in Section 1, a key challenge for AMT is the lack of a general and sizable dataset for training and evaluation. Annotating a ground-truth transcription for polyphonic music is highly time-consuming and requires significant expertise. One way to address this challenge is to generate chord and note progressions synthetically. This can be accomplished by manually creating a set of midi files using libraries that provide support for MIDI generation. The generated composition can be beneficial for experimentation, testing, and for automatic music transcription of piano songs.

To generate a synthesized MIDI piano song, notes from $C2$-$C8$ have been selected and played in sequence. Also, a set of common chords were chosen to build a polyphonic data set, including major triads, minor triads, dominant $7^{th}$, minor $7^{th}$, major $7^{th}$, $5^{th}$ chords, major $6^{th}$, minor $6^{th}$, diminished triad, suspended $2^{nd}$, suspended $4^{th}$, augmented triad, and augmented $7^{th}$ chords. As a result, the MIDI piano song contains a diverse set of individual notes, as well as multi-note chord progressions with two, three, and four note chords.

### 4.1. Pre-processing

The pre-processing stage involves extracting features from the input data (piano music data) and preparing them for training. In the case of music transcription, one common feature extraction method is the CQT. The audio file of the piano music is stored in WAV file format, and these files hold a 1-dimensional array of numbers that represent sound pressure over time. During the data pre-processing stage, the input pre-processing script takes a WAV file and splits it into smaller audio frames. Each audio frame consists of 2756 samples, obtained at a sampling rate of $44.1kHz$, representing 0.0625 seconds of a musical piece. Next, all the frames are converted into the frequency-time domain using the CQT method.

For generating the target training data, the accompanying MIDI files of the WAV piano musical files are parsed using the prettymidi library[1]. First, we start by determining the duration of the piano midi song and then compute the total frames that we can get based on a frame duration (which is 0.0625 seconds in this experiment). We can get the duration of a song by checking the end time of the last midi note event. Then, we divide the song duration by the frame duration to get the total number of frames. This is important to match each frame of the audio chunk to the corresponding MIDI file chunk. The column of each frame represents a vector that has a dimension of 72 corresponding to the possible pitches of a piano key from C2-C8. Then, we can loop over each midi event and assign notes that are active on each frame. The training files are then saved separately. Once these representations are obtained and stored in files, our network can directly process them and learn from them.

### 4.2. Silent Frames

The energy of the sound generated by a piano notes decays over time as time passes after the string has been struck [16]. This characteristic becomes more noticeable as we move from lower to higher pitch notes. The decreasing energy of a musical note causes several problems for the music transcription task. The main problem is the difficulty of determining the offset time if a note sound diminishes faster. One solution to this problem is to determine a threshold of energy of frames, which will determine when to consider a frame as silent and remove them from the training data. In this experiment, a standard deviation threshold of 0.00001 has been chosen and frames below this threshold are ignored from the training and testing data.

---

[1] https://craffel.github.io/pretty-midi/

## 5. Experimental Results

### 5.1. Metrics

In the experiment, a frame-wise accuracy measure was used to evaluate the performance of the model [17]. If the predicted labels for a specific frame completely match the true labels, then the frame is considered to be correctly predicted, resulting in a subset accuracy of 1.0. However, if there are any differences between the predicted and true labels, the subset accuracy is 0.0.

$$accuracy(y, \hat{y}) = \frac{1}{n_{samples}} \sum_{i=0}^{n_{samples}-1} 1(\hat{y}_i = y_i) \quad (1)$$

where $\hat{y}_i$ is the predicted value of the $i$-th frame and $y_i$ is the corresponding true value and $1(x)$ is the indicator function.

### 5.2. Experimental Set-up

The training experiments were performed using a workstation equipped with a 40 cores Intel(R) Xeon(R) CPU E5-2698 v4 @ 2.20GHz processor. The workstation was also equipped with 252 GB of RAM and a Tesla V100-DGXS-32GB GPU for processing.

A total of 830 piano songs, each lasting 3 minutes have been generated. Out of these, 80 percent is used for training and 20 percent is used for validation. Additionally, a dataset of 166 piano songs (the same size as the validation set) was generated with uniform distribution and used as a test set to compare the performance of the models.

### 5.3. Model Performance

Figure 3 depicts the evolution of the training and validation accuracy while training the holistic model. A batch size of 64 was used to train the models. From the training plot, we can understand that convergence is achieved after 70 epochs.

Additionally, another version of *Holistic V1*, referred to as *Holistic V2*, was trained by removing silent frames (refer 4.2). The training result of this model is depicted in Fig 4.

Table 1 summarizes the performance of the models. The models were tested on their inference time, accuracy, and time taken to train. The inference time is the time it takes for the models to make a prediction once they are given the input. The inference duration is measured based on the time taken by the model to make a prediction for a 2-minutes long piano song.

### 5.4. Model Prediction

The Figure 5 shows example model predictions from structured models and holistic models visualized through a piano roll representation. The green dots represent the false negatives, indicating the notes and chords the models were not able to predict correctly, and the red dots represent false positive predictions. The dark green dots and segments represent true positives where the models make the correct predictions. Overall, the model performed well on the prediction task, as evident from the high number of true positive predictions. However, as shown in the piano roll visualization, the models made some false positive and false negative predictions.

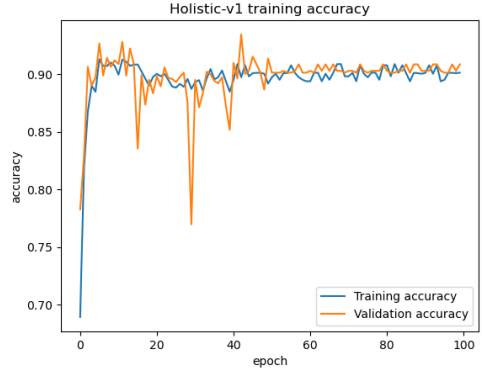

**Figure 3:** Training plot of the holistic models

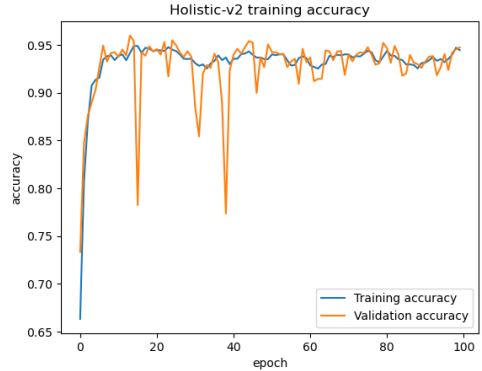

**Figure 4:** Training plot of the holistic-v2 models

| Model | TT(minutes) | IT(msec) | Acc(%) |
|---|---|---|---|
| Holistic - v1 | 793.6 | 660 | 88.2 |
| Holistic - v2 | 720.2 | 664 | 92.5 |

**Table 1**

Model result summary, including training time (TT), inference time (IT) and accuracy (Acc).

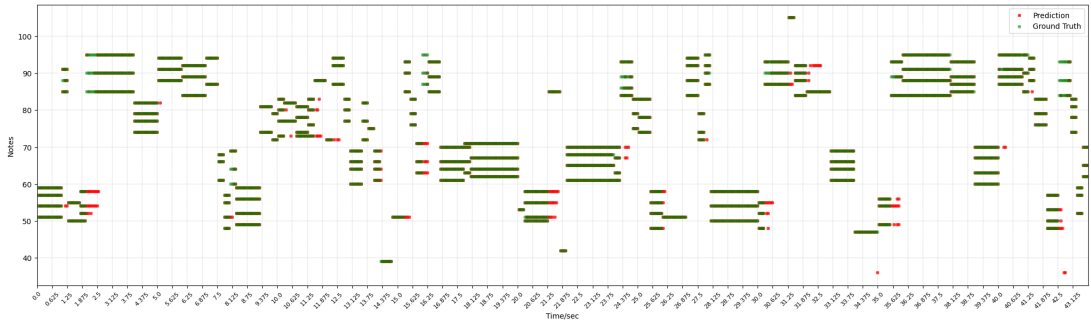

**Figure 5:** Example prediction - piano roll representation.

One interesting but expected finding is that most of the false positive predictions occur during the transition from one chord/note to another. This is because the frames close to the transition of notes and chords could contain the frequency of notes/chords from both sides, leading to a potential confusion and notes/chords from both side being predicted. Moreover, it is observed that these false positive predictions are often notes from the previous notes/chord that are still audible. This is expected since when we release a note on the piano, the sound might not end immediately and the note will be heard for a short period of time, leading to the prediction of the same note/chord after the note/chord ended.

## 6. Conclusions and Future Works

We have presented an approach to tackle the problem of AMT for polyphonic piano music using CNN and CQT, showing experimental results about the achieved accuracy and performance.

There are many ways to improve this AMT system. This work and accompanying experimentation are based on fully synthesized audio data. Extending this approach using real piano audio data for the training and/or testing phases could provide more diverse training examples for the AMT system, leading to a more realistic measurement of its achievable performance in real-world scenarios.

Now that we have baseline results using CNN, a next step would be to explore other models, such as RNN, that might capture the longer dependencies between notes and chords. By exploiting the temporal nature of music, RNNs can potentially improve the accuracy of the proposed AMT system.

## References

[1] W. contributors, Transcription (music), https://en.wikipedia.org/wiki/Transcription_(music), 2022.

[2] P. Smaragdis, J. Brown, Non-negative matrix factorization for polyphonic music transcription, 2003 IEEE Workshop on Applications of Signal Processing to Audio and Acoustics (IEEE Cat. No.03TH8684) (2003) 177–180.

[3] S. Abdallah, M. Plumbley, Polyphonic music transcription by non-negative sparse coding of power spectra, in: International Society for Music Information Retrieval Conference, 2004.

[4] A. Cont, Realtime multiple pitch observation using sparse non-negative constraints, in: ISMIR 2006, 7th International Conference on Music Information Retrieval, Victoria, Canada, 8-12 October 2006, Proceedings, 2006, pp. 206–211.

[5] A. Zalani, A. Mittal, Polyphonic music transcription a deep learning approach, https://cse.iitk.ac.in/users/cs365/2014/_submissions/aniruddh/project/report.pdf, 2014.

[6] D. E. Rumelhart, G. E. Hinton, R. J. Williams, Learning internal representations by error propagation, in: D. E. Rumelhart, J. L. Mcclelland (Eds.), Parallel Distributed Processing: Explorations in the Microstructure of Cognition, Volume 1: Foundations, MIT Press, Cambridge, MA, 1986, pp. 318–362.

[7] G. E. Hinton, A practical guide to training restricted boltzmannmachines, in: Proceedings of Neural Networks: Tricks of the Trade (2nd ed.), 2012, pp. 599–619.

[8] V. N. Vapnik, Introduction: Four Periods in the Research of the Learning Problem, Springer New York, New York, NY, 2000, pp. 1–15. URL: https://doi.org/10.1007/978-1-4757-3264-1_1. doi:10.1007/978-1-4757-3264-1_1.

[9] L. E. Baum, T. Petrie, G. Soules, N. Weiss, A

maximization technique occurring in the statistical analysis of probabilistic functions of markov chains, The Annals of Mathematical Statistics 41 (1970) 164–171. URL: http://www.jstor.org/stable/2239727.

[10] S. Sigtia, E. Benetos, S. Dixon, An end-to-end neural network for polyphonic piano music transcription, https://arxiv.org/abs/1508.01774, 2015. doi:10.48550/ARXIV.1508.01774.

[11] B. Uria, M.-A. Côté, K. Gregor, I. Murray, H. Larochelle, Neural autoregressive distribution estimation, Journal of Machine Learning Research 17 (2016) 1–37. URL: http://jmlr.org/papers/v17/16-272.html.

[12] R. Kelz, M. Dorfer, F. Korzeniowski, S. Böck, A. Arzt, G. Widmer, On the potential of simple framewise approaches to piano transcription, https://arxiv.org/abs/1612.05153, 2016. doi:10.48550/ARXIV.1612.05153.

[13] V. Emiya, N. Bertin, B. David, R. Badeau, Maps - a piano database for multipitch estimation and automatic transcription of music, 2010.

[14] C. Hawthorne, E. Elsen, J. Song, A. Roberts, I. Simon, C. Raffel, J. Engel, S. Oore, D. Eck, Onsets and frames: Dual-objective piano transcription, https://arxiv.org/abs/1710.11153, 2017. doi:10.48550/ARXIV.1710.11153.

[15] N. Sharma, V. Jain, A. Mishra, An analysis of convolutional neural networks for image classification, Procedia Computer Science 132 (2018) 377–384. URL: https://www.sciencedirect.com/science/article/pii/S1877050918309335. doi:https://doi.org/10.1016/j.procs.2018.05.198, international Conference on Computational Intelligence and Data Science.

[16] T.-L. Cheng, S. Dixon, M. Mauch, Modelling the decay of piano sounds, International Conference on Acoustics, Speech, and Signal Processing (2015). doi:10.1109/icassp.2015.7178038.

[17] scikit-learn accuracy score, https://scikit-learn.org/stable/modules/model_evaluation.html#accuracy-score, 2023. Accessed: 2023-02-30.